\documentclass[twocolumn,trackchanges]{aastex631}

\usepackage{mathptmx}
\usepackage[T1]{fontenc}
\usepackage{ae,aecompl}
\usepackage{graphicx}	
\usepackage{amsmath}	
\usepackage{amssymb}	

\begin{document}

\title{Electrostatic Repulsion of Dust from Planetary Surfaces}

\author[0000-0001-8486-4743]{F. Chioma Onyeagusi}
\affiliation{University of Duisburg-Essen, Faculty of Physics,
Lotharstr. 1-21, 47057 Duisburg, Germany}

\author[0000-0002-0269-2763]{Felix Jungmann}
\affiliation{University of Duisburg-Essen, Faculty of Physics,
Lotharstr. 1-21, 47057 Duisburg, Germany}

\author[0000-0003-4468-4937]{Jens Teiser}
\affiliation{University of Duisburg-Essen, Faculty of Physics,
Lotharstr. 1-21, 47057 Duisburg, Germany}

\author[0000-0002-7962-4961]{Gerhard Wurm}
\affiliation{University of Duisburg-Essen, Faculty of Physics,
Lotharstr. 1-21, 47057 Duisburg, Germany}

\begin{abstract}
    Surfaces of planetary bodies can have strong electric fields, subjecting conductive grains to repulsive electrostatic forces. This has been proposed as mechanism to eject grains from the ground. To quantify this process, we study mm-sized basalt aggregates consisting of micrometer constituents exposed to an electric field in drop tower experiments. The dust aggregates acquire high charges on sub-second timescales while sticking to the electrodes according to the field polarity. Charging at the electrodes results in a repulsive (lifting) force and continues until repulsion overcomes adhesion and particles are lifted, moving towards the opposite electrode. Some aggregates remain attached, which is consistent with a maximum charge limit being reached, providing an electrostatic force too small to counteract adhesion. All observations are in agreement with a model of moderately conductive grains with a small but varying number of adhesive contacts to the electrodes. This supports the idea that on planetary surfaces with atmospheres, electrostatic repulsion can significantly contribute to airborne dust and sand, i.e. decrease the threshold wind speed that is required for saltation and increase the particle flux as suggested before.
\end{abstract}


\section{Introduction}

During eolian activity on Earth, electric fields on the order of 100 kV/m have been reported close to the ground \citep{Schmidt1998, Zheng2013, Zhang2020}. Similar values have been deduced from hovering dust seen as lunar horizon glow on the Moon \citep{Rennilson1974}. Significant electrostatic forces inevitably act on the soil in such a strong field.
On one side, electric fields might influence aggregation of ejected dust grains simply by inducing dipoles which attract each other \citep{Jungmann2022}. On the other side, the fields might be particularly important in the context of lifting grains from the surface in the first place \citep{Renno2008, Kimura2022, Hirata2022, Wang2016}.

To quantify particle lifting in electric fields, \citet{Kok_and_Renno_2006} placed natural soil samples in an electric field and measured the lifted sample mass at certain field strengths. They found that 100 µm grains were lifted by fields starting at about 180 kV/m. With increasing electric field they also observed a steep incline in lifted mass. Along the same line of increasing the mass budget of airborne dust, \citet{Esposito2016} showed through field campaign measurements that electrostatics can enhance the amount of particles emitted into the atmosphere even by a factor of 10 above regular gas drag in a dust storm. 

The most commonly assumed lifting mechanism related to electric fields requires conductive grains. Such grains would charge with the respective polarity of the ground, generating a repulsive (lifting) force. In agreement with this, \citet{Holstein2012} showed that the threshold friction velocity in wind tunnel experiments indeed decreased for grains (glass, quartz, copper) on a conductive surface, while the threshold friction velocity necessary for saltation increased for grains on an insulating surface.
This electrostatic repulsion mechanism of conducting grains has only recently been suggested for micrometer particles on other celestial bodies like asteroid Phaeton or Saturn's rings \citep{Kimura2022, Hirata2022}. Especially the latter works require additional ways to reduce the sticking forces that counteract the repulsive force, but otherwise build on the same electrostatic repulsion.

To verify this hypotheses of repulsion supported by conduction and to quantify conditions for particle ejection with respect to cohesion, we study particle lifting on a microphysical level. In view of gravity being a dominant force on Earth, we carried out drop tower experiments. During the microgravity phase, dust aggregates collide with or stick to the electrodes of the electric field capacitor. In the absence of gravity, it is only a matter of cohesion and electrostatic force whether particles are released or not. Therefore, the drop tower experiments allow us to quantify the grains' charge, sticking forces and (re)-charging timescales in detail.

\section{Experiments}

\begin{figure}
    \centering
    \includegraphics[width= \columnwidth]{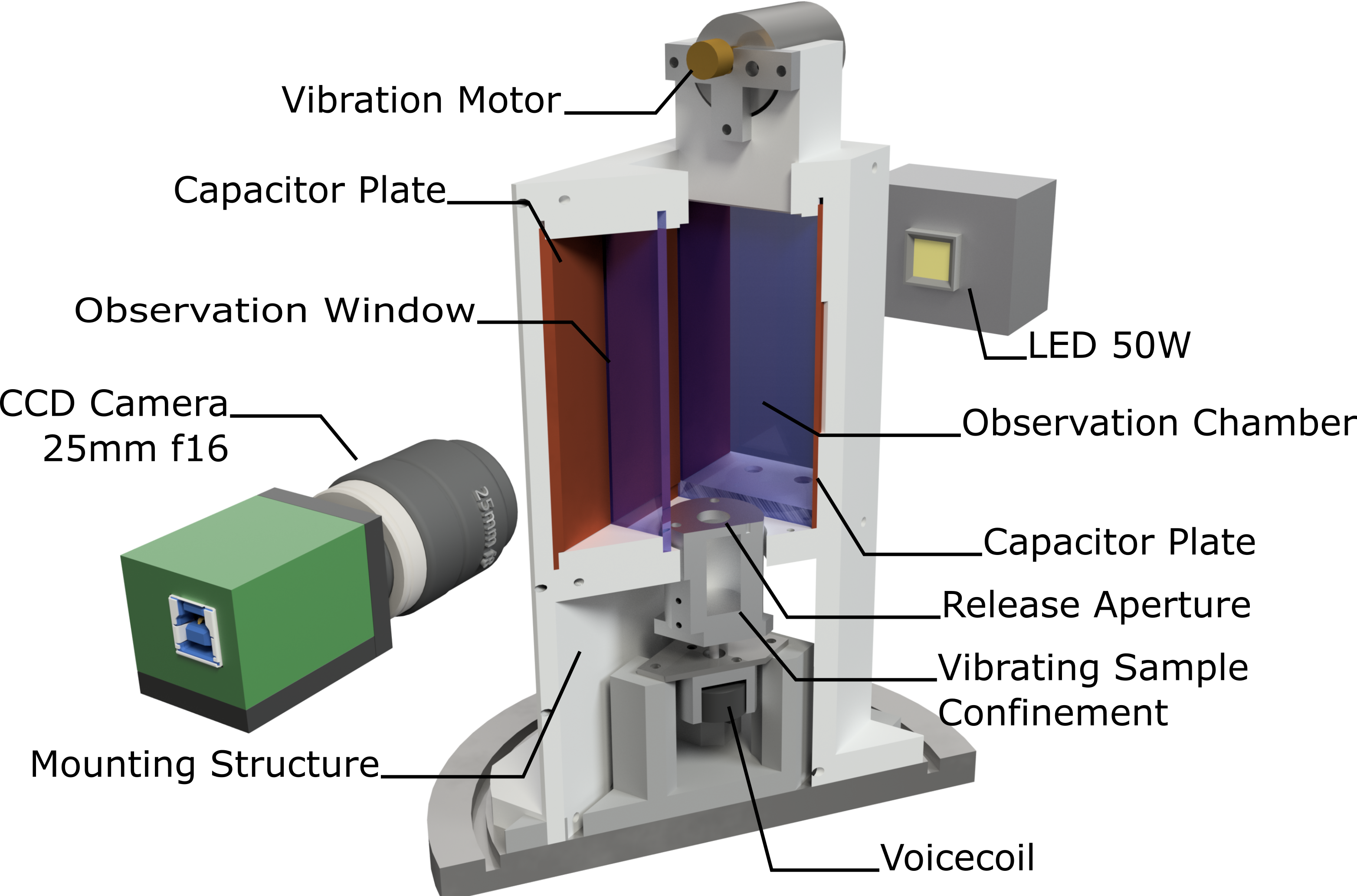}
    \caption{Schematics of the experiment setup (from \citet{Steinpilz2020a})}
    \label{fig:setup}
\end{figure}

\begin{figure}
    \centering
    \includegraphics[width= \columnwidth]{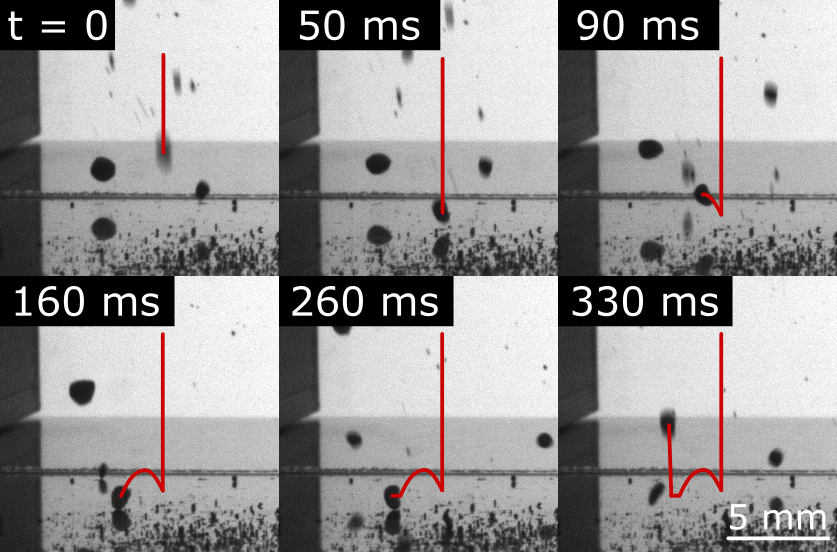}
    \caption{Time series of aggregates observed during the drop tower experiments at one of the copper electrodes (bottom). Particles appear twice due to the copper's high reflectivity. Upon sticking, small aggregates remain stuck while large aggregates recharge and are repelled. Here, the trajectory of one aggregate is marked showing a rebound, sticking and repulsion after recharging at the electrode within the electric field. }
    \label{fig:grainsatflight}
\end{figure}

\begin{figure}
    \centering
    \includegraphics[width= \columnwidth]{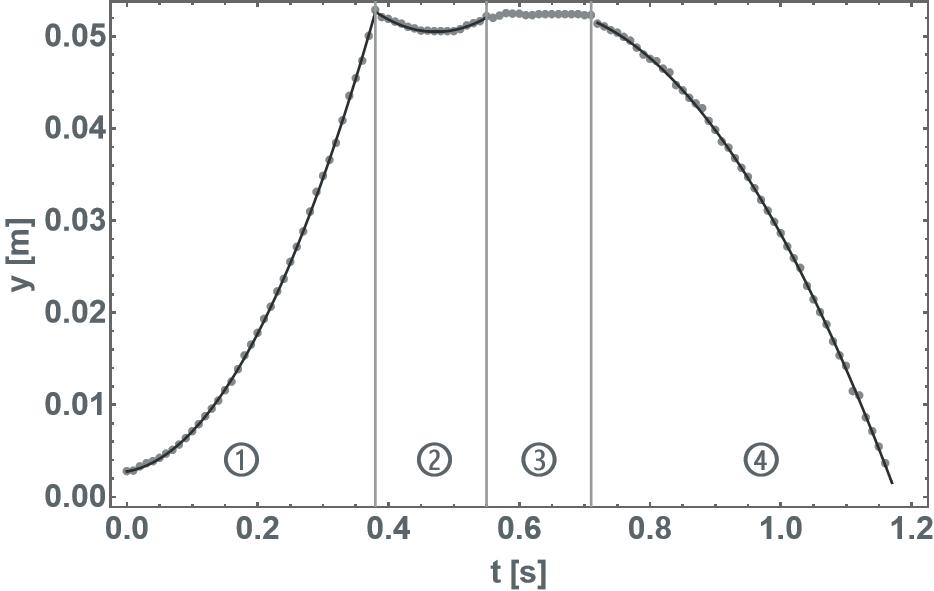}
    \caption{Distance of an aggregate to an electrode (located at top horizontal) over time. Four different regions are marked; 1: Accelerated approach due to Coulomb attraction; 2: Approach after bouncing with same polarity; 3: Sticking on electrode; 4: Accelerated repulsion due to Coulomb force with opposite polarity; Parabolic fits are plotted for regions 1, 2, and 4.}
    \label{fig:bouncstickfly}
\end{figure}

Each experiment consists of a preparation phase on ground and the microgravity phase after the setup is launched in the drop tower in Bremen.
A sketch of the drop tower setup taken from \citet{Steinpilz2020a} is shown in 
fig. \ref{fig:setup}.
As soon as microgravity sets in, the sample confinement starts shaking, injecting the aggregates through its opening into the experiment chamber, which also operates as capacitor. The particle motion within the electric field is observed by bright field imaging, where particles appear dark in front of a bright background in the video data. The resulting videos are tilted by 90°. An example of aggregates in microgravity is shown in fig. \ref{fig:grainsatflight}. 
After launch, microgravity lasts about 9~s with residual gravity being on the order of $10^{-5} ~ \rm g$. This is insignificant for the particle motion in our context. We used air at 1~bar as ambient atmosphere in the current setting. The humidity within the experiment chamber averaged at 29.5 \% at the time of launch.
After the sample is released, the charged grains are accelerated by the electric field of the capacitor. Observing the motion of the charged grains allows the determination of the charge distribution on aggregates right after injection, before and after bouncing collisions and after sticking and repulsion at an electrode. These different phases of particle motion are visualized in fig. \ref{fig:bouncstickfly}. It shows the distance of an aggregate to one of the electrodes (located at the top horizontal axis) over time. The aggregate is accelerated towards the electrode and bounces off, keeping the same polarity. It collides again, but this time it sticks until it is recharged sufficiently to be ejected against cohesion, now having a different polarity.

\subsection{Particle sample}
\begin{figure}
    \centering
    \includegraphics[width= \columnwidth]{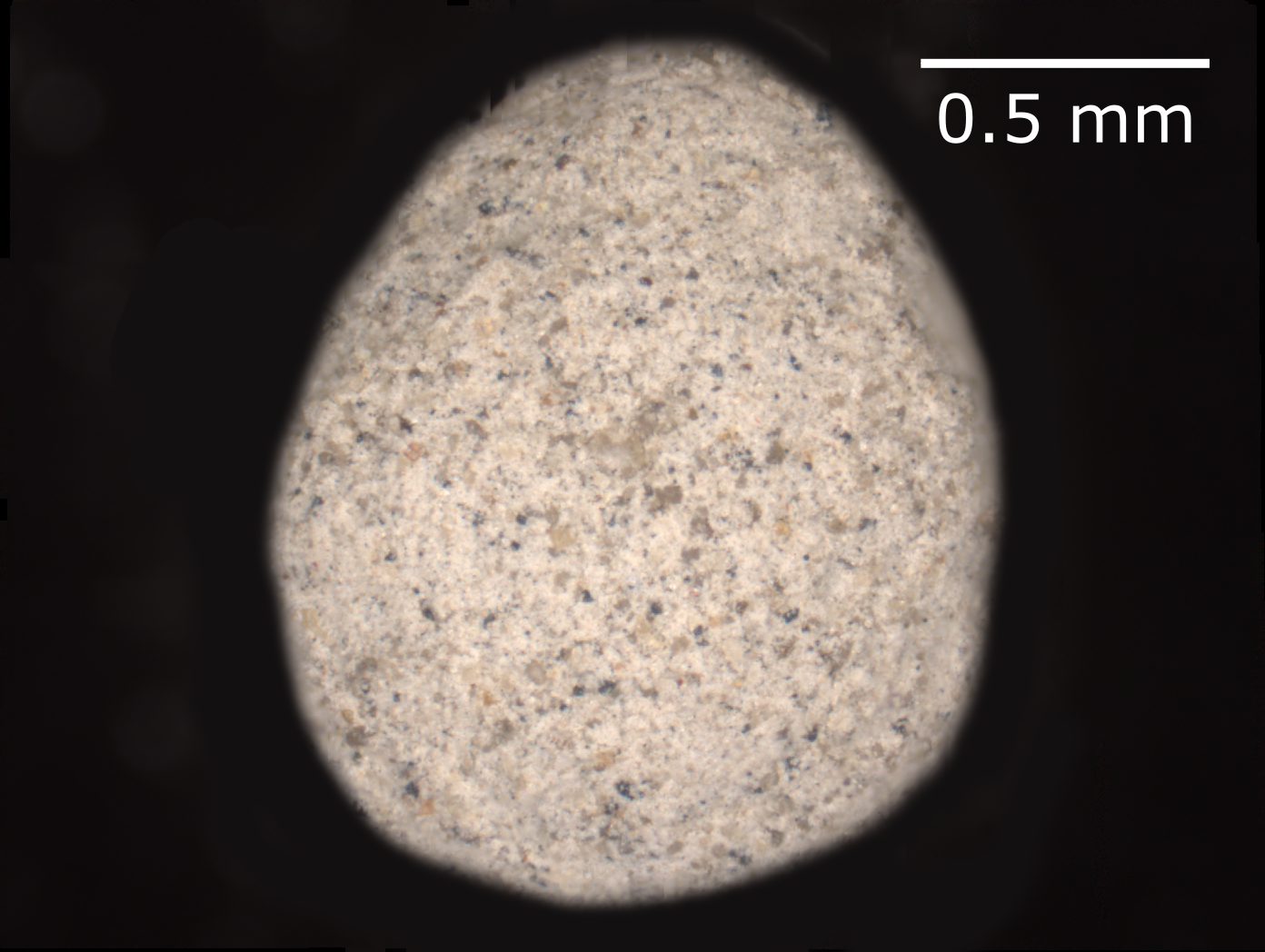}
    \caption{Composite microscope image of a dust aggregate. Individual dust grains of the compact aggregate are visible as granular texture.}
    \label{fig:dustaggregate}
\end{figure}

\begin{figure}
    \centering
    \includegraphics[width= \columnwidth]{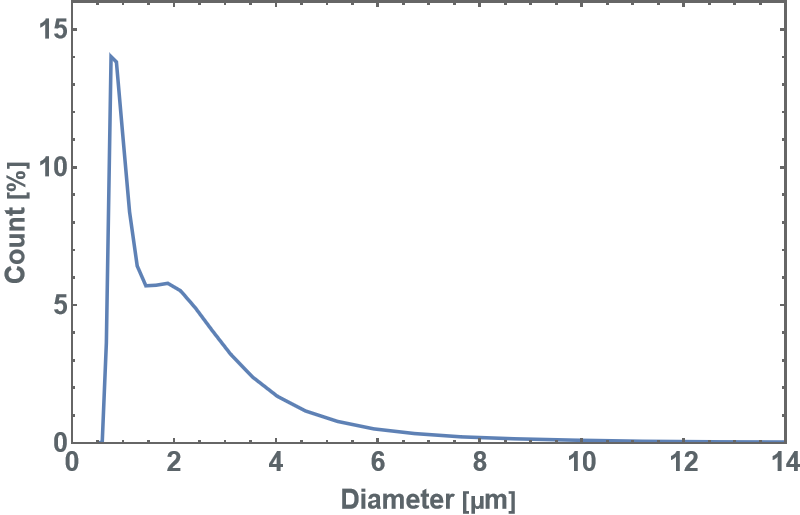}
    \caption{Number-size distribution of the milled dust grains that were used to form aggregates as measured by a Mastersizer 3000 by means of wet dispersion.}
    \label{fig:sizedistributiondust}
\end{figure}

In earlier experiments we used monolithic glass or basalt beads with the same setup but a different scientific focus \citep{Steinpilz2019, Jungmann2018, Jungmann2021b}. This paper reports the first experiments with dust aggregates. 
As far as ground preparation goes, the aggregates were created in the laboratory by strong vibrations of a dust reservoir. During this procedure, individual dust grains (micrometer-sized) stick together until large dust aggregates are formed, compacted and then only bounce off each other \citep{Zsom2010, Kelling2014, Kruss2016, Weidling2009}. This procedure leads to the formation of mostly spherical or elliptical aggregates in a limited size range on the order of 1 mm and with average volume filling factors of about 0.52 (assuming a density of $2700 ~ \rm kg/m^3$ for the constituent grains). 
The preparation process is carried out days to weeks before the drop tower campaign.
A microscope image of such a dust aggregates is shown in fig. \ref{fig:dustaggregate}. The image was generated using a stack of images with varying focal lengths.
The underlying dust was produced by milling larger basalt particles ($\le$~200~µm) to a final grain size of a few µm using a planetary ball mill (PM 100 CM). The milled µm-grains exhibit an irregular shape. A number-size distribution of the milled dust grains is shown in fig. \ref{fig:sizedistributiondust}. 

The pre-produced dust aggregates are placed in a sample confinement as shown in fig. \ref{fig:setup} and are vibrated for about 15 minutes just prior to launch into microgravity. In earlier experiments, these vibrations led to tribocharging. This was attempted here as well, in order to see how pre-charged grains behave upon collisions within the capacitor. The walls of the reservoir are coated with the same basaltic dust to avoid a material bias for contact charging. Vibrations occur at lower amplitudes compared to the preparation process and are stopped at maximum one minute before launch; this was not quantified further.

The particle trajectories are traced by tracking the center of mass of the aggregates using ImageJ \citep{Rasband1997}. The mass is deduced from the average cross section by assuming a sphere of equivalent cross section and using the given average filling factor. 
The y-trajectories towards the electrode plates are approximated as parabolic functions $(a/2) t^2$ with the acceleration $a$. With this assumption, the charge on a grain is  given by 
\begin{equation}
    q = \frac{m a}{E} = \frac{m a d}{U}
\end{equation}
with the electric field $E = U/d$ given by the applied voltage $U$ and the distance $d$ between the electrodes.

We note that the trajectories are well approximated by parabolas, so gas drag is not important. In any case, for a typical grain with a radius $r = 0.5 $\,mm and a velocity of  $v = 0.01$\,m/s and assuming a dynamic viscosity of $\eta = 1.8 \cdot 10^{-5}$\,Pa\,s for air, the Stokes drag amounts to  $F_s=6 \pi \eta r v = 2 \cdot 10^{-9}$\,N. Compared to the Coulomb force on a charge of $q = 1$\,pC in an electric field $E = 1.6 \cdot 10^5 $\,V/m resulting in $F_C = 2 \cdot 10^{-7}$\,N the Stokes drag is 2 orders of magnitude smaller.

\section{Results}

As result we get different charge distributions at various times, depending on the contact history and contact duration with the electrodes. As the aggregates have a certain size variation and the grains' charge might depend on size, we split the data in two fractions being either smaller or larger than 1\,mm. The smallest tracked aggregates have an average diameter of 0.4\,mm, the largest 2.2\,mm. The database is not large enough to study a systematic size distribution in more detail, but the chosen separation into a small and a large fraction already demonstrates trends and provides plausibility checks.

\subsection{Pre-charge}

\begin{figure}
    \centering
    \includegraphics[width= \columnwidth]{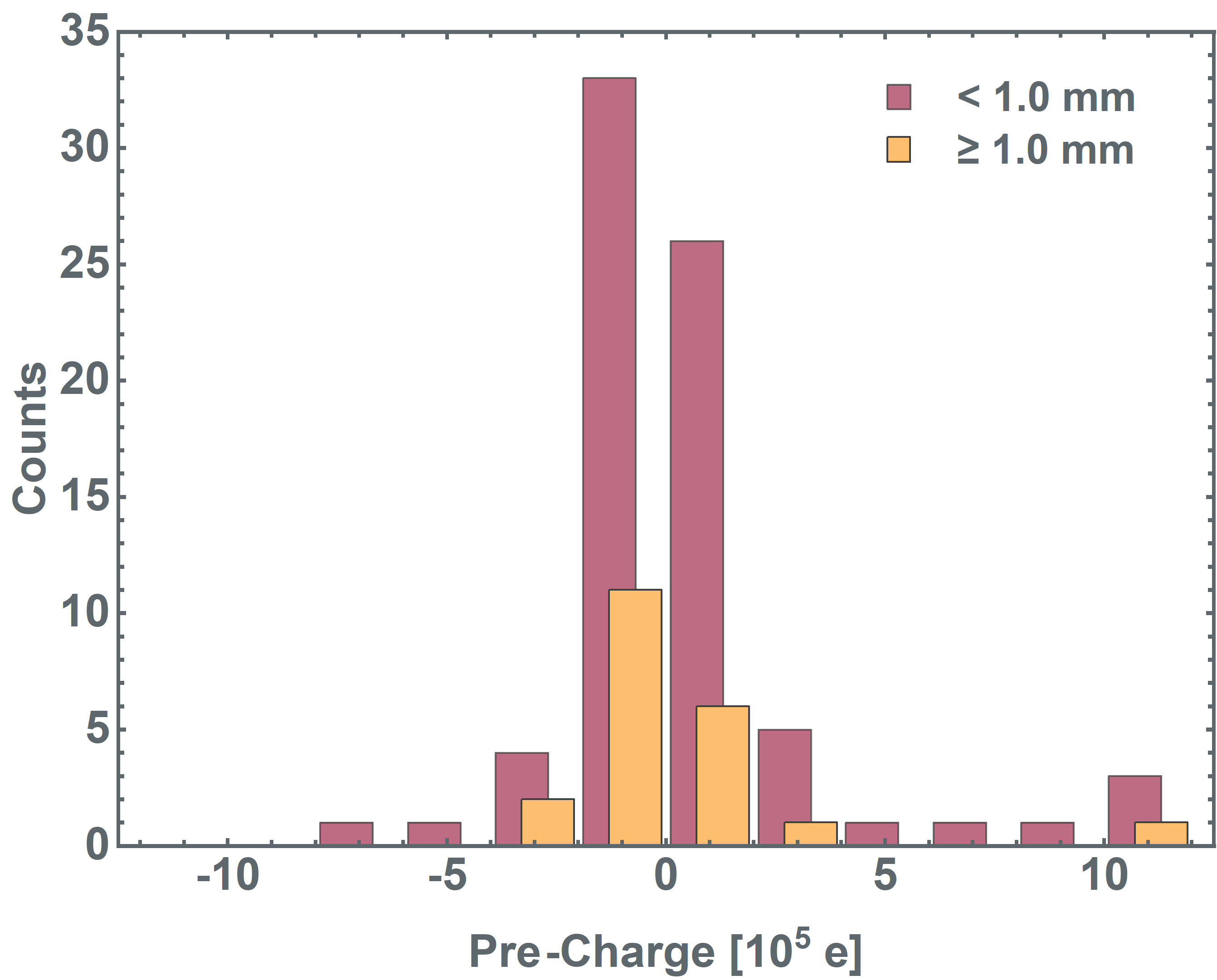}
    \caption{Initial charge distribution measured after grains entered the capacitor.}
    \label{fig:initial}
\end{figure}

The first charge distribution refers to the grains right after entering the capacitor and before hitting and potentially sticking onto or recharging at an electrode. This distribution is shown in fig. \ref{fig:initial}.
The charge distribution is centered around zero. As expected, there is no polarity bias. However, in earlier experiments with monolithic (non conducting) particles, mm-grains charged up to values of a few $10^8 \, \rm e$ (e.g. 0.4\,mm glass beads in \citet{Jungmann2022x}). In comparison, the 1\,mm dust aggregates exhibit about 2 orders of magnitude less charge. This already implies that collisional charging, in this case, is very ineffective and / or discharge by conduction or other means is very effective during the short time period prior to launch, when grains are not vibrated.

\subsection{Recharging upon sticking}

\begin{figure}
    \centering
    \includegraphics[width= \columnwidth]{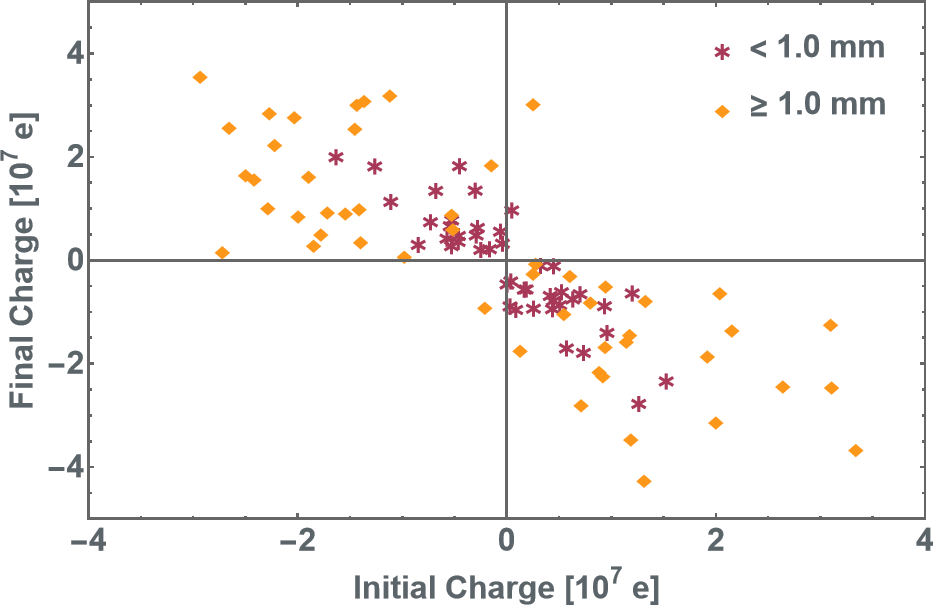}
    \caption{Recharging at the electrodes in sticking collisions. Negatively charged grains recharge positively at the positive electrode and vice versa.}
    \label{fig:recharging}
\end{figure}

As first category of electric charging in the context of collisions with the electrodes, we show cases where aggregates stick to the copper plates for fractions of a second before they are repelled again. Fig. \ref{fig:recharging} shows the charges for aggregates prior to sticking and after ejection.
Positive grains drawn to and impacting the negative electrode recharge negatively and vice versa. The lower left and upper right quadrant of the graph are essentially empty. Evidently, the grains recharge with the polarity expected for conductors at the respective electrode.  

\begin{figure}
    \centering
    \includegraphics[width= \columnwidth]{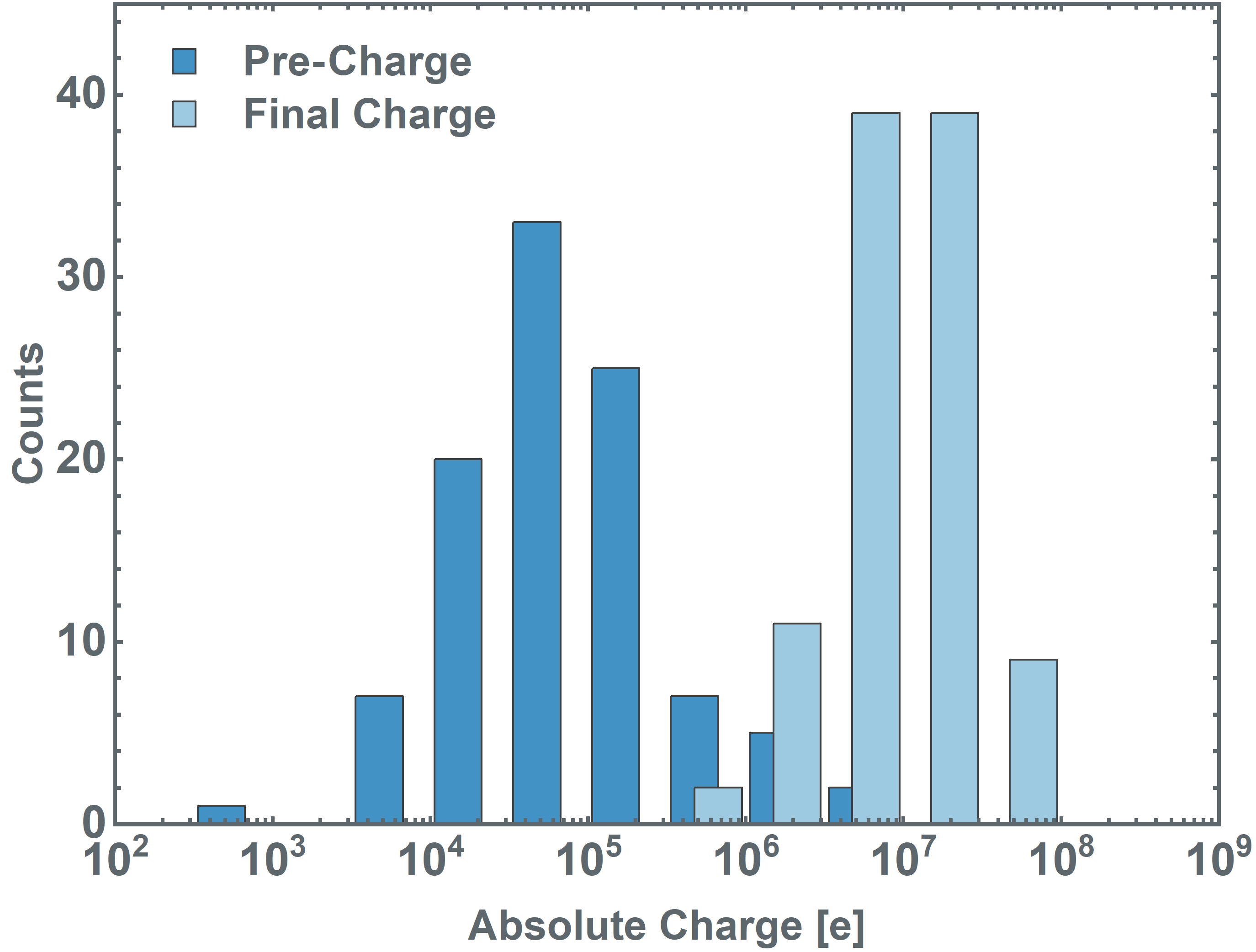}
    \caption{Comparison of the pre-charge on grains and the charge gained during recharge at the electrodes.}
    \label{fig:compare}
\end{figure}

There is no single, specific value to which the grains charge before they are ejected again, but there is a wide spread charge distribution. On the upper end, it can reach several $10^7 \, \rm e$, which is about 2 orders of magnitudes larger than the pre-charge, as seen in comparison in fig. \ref{fig:compare}. 

\begin{figure}
    \centering
    \includegraphics[width= \columnwidth]{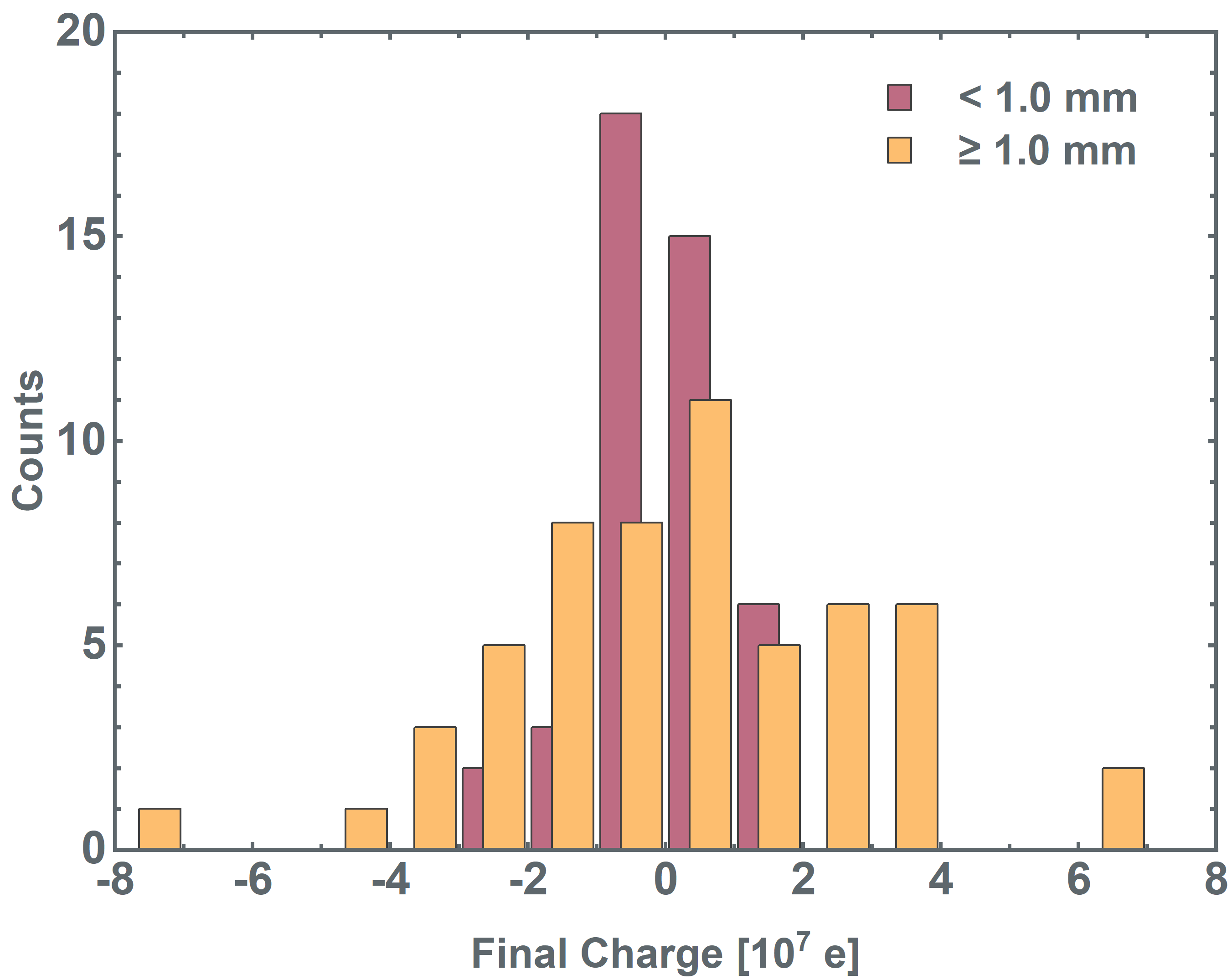}
    \caption{Charge distribution after recharging on electrodes once grains are ejected after sticking.}
    \label{fig:recharge_histo}
\end{figure}

Besides, larger grains typically have higher absolute charges. This is also true for the charge prior to collisions, noting that this initial charge before a collision is typically not the pre-charge from the previous section. Grains bounce back and forth several times between the electrodes, so that the initial charge is set by the last recharge at an electrode. Accordingly, the range of charges before and after a sticking contact with an electrode are consistently on the same order of magnitude. The exact amount of charge following the same aggregate can be different from recharge to recharge. The charge distribution of ejected aggregates is shown in fig. \ref{fig:recharge_histo}. 

\begin{figure}
    \centering
    \includegraphics[width= \columnwidth]{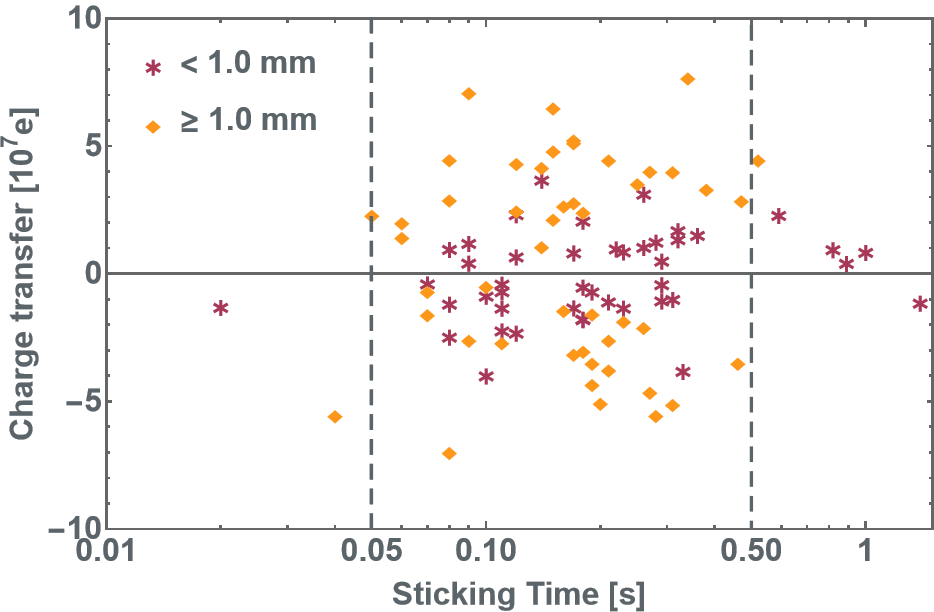}
    \caption{Charge transfer in a sticking event depending on sticking time. Dashed lines roughly mark the range of times during which ejections of the large grains occur. Ejection of smaller grains is slightly shifted to larger time intervals.}
    \label{fig:timing1}
\end{figure}

The timescale of recharging and, therefore, of sticking might also be of importance. Charge transfer in dependence of sticking time is depicted in fig. \ref{fig:timing1}.
There is no systematic time dependence of the charge transfer for a given aggregate size, but for the large aggregates repulsion seems restricted to a time interval from about 0.05 to 0.5 s after a particle reaches the copper plate. For smaller aggregates this interval appears to be shifted slightly towards larger sticking times. In any case, charging is not immediate and not ongoing for longer timescales. Aggregates smaller than 0.2\,mm are usually not rejected from the electrodes at all. This also applies to remnants of aggregates that are at times being shed upon collision with an electrode. In those cases, electrostatic forces are not strong enough to eject the small aggregates, while ejectable aggregates need a certain time to build up charge until reaching the final charge density.

\subsection{Recharging upon bouncing}

\begin{figure}
    \centering
    \includegraphics[width= \columnwidth]{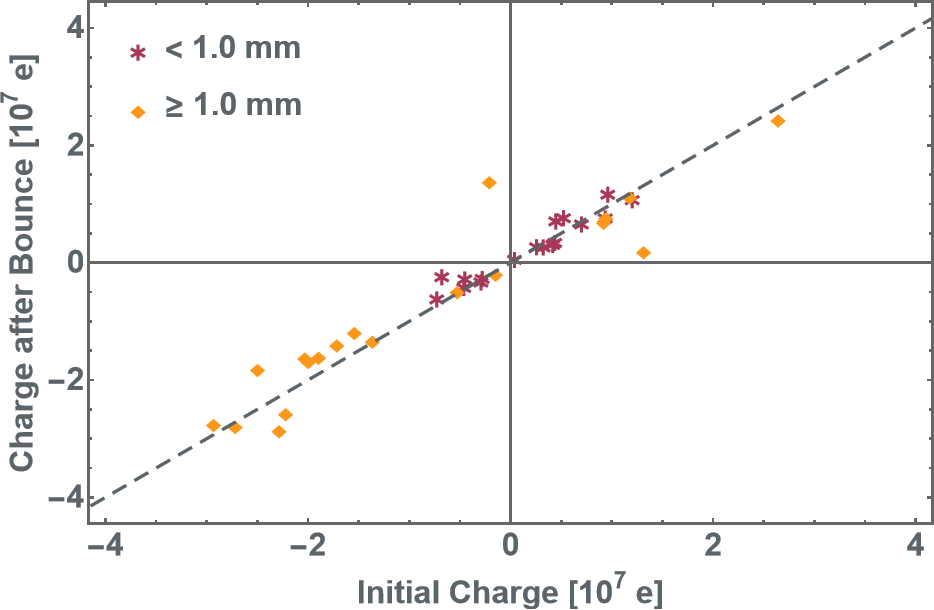}
    \caption{Charge on aggregates before and after a bouncing collision. The dashed line marks charge conservation.}
    \label{fig:bounce}
\end{figure}

The second category of electrode collisions considers bouncing collisions, where aggregates hit an electrode and bounce off again. Typically, the particles are accelerated towards the same electrode they bounced off of, resulting in a parabolic motion, as seen in fig. \ref{fig:bouncstickfly} section 2. This already suggests that the aggregates do not change their polarity during the short contact. Here, contact times are on sub-ms timescales \citep{Kelling2014}. This is well below the timescale at which aggregates are rejected again after sticking.  
The charge of grains before and after a collision are shown in fig. \ref{fig:bounce}. 

This is the analog to fig. \ref{fig:recharging}.
In contrast, however, only little charge is transferred in these collisions on the uncertainty level of the measurements of a few \%.

\section{Repulsion model}

As particles recharge, they experience a force repelling them from the electrode to which they stick to. The easiest assumption is that particles recharge to a point when the repelling Coulomb force equals the sticking or adhesive force as these are the two dominating forces acting on a dust aggregate in the absence of gravity. 

\textbf{Adhesion:} As far as sticking goes for dust aggregates, contact in a bouncing or sticking collision does not occur in an elastic and rigid manner; dust grains can locally rearrange within the aggregate \citep{Kruss2016,Jankowski2012}. Therefore, the number of contacts is not fixed, but can be anything larger than 1.
There is evidence in the data from aggregates gently bending down after contact, kind of swaying in the field, which implies that not more than two effective contacts have been made. However, not all aggregates behave this way and with an irregular surface larger numbers of contacts are possible as well. 
Quantitatively, the adhesive force of $N$ contacts of dust grains with a radius $r_d$ and a flat surface is
\begin{equation}
F_{ad} = 3 \pi \cdot \gamma \cdot r_d \cdot N
\end{equation}
according to the JKR model \citep{Johnson1971}. The surface energy $\gamma$, in principle, is a well defined quantity for a given material. \citet{Pillich2021} measured the surface energy of basalt aggregates of the same material with a similar filling factor and constituent grain size. They determined a value of $\gamma =0.02 ~ \rm J/m^2$. In our case, the metal electrode needs to be taken into account as well. If the effective surface energy between grain and electrode was smaller, aggregates would be lifted easier. If the surface energy was much larger, aggregates would not be lifted or regularly lose particles as cohesion between grains would be broken easier. As we (sometimes) observe this, we assume the typical silicate surface energy to be a reasonable assumption to make.


\textbf{Electrostatic repulsion:} 
The charge density (charge $q$ over area $A$) on the conductive surface of a capacitor is
\begin{equation}
    q/A = \epsilon_0 E.
\end{equation}
A particle in contact, charging over time, should at maximum gather a similar charge density. If we take $A = 4 \pi r^2$ as the particle's surface, the maximum charge on the grain is
\begin{equation}
    q_{max} = 4 \pi \epsilon_0 E r^2 = \frac{4 \pi \epsilon_0 U r^2}{d}. 
    \label{qabs}
\end{equation}
Within a small factor this results in an electrostatic repulsive force $F_{el, max} = q_{max} E$ similar to the force used by \citet{Kok_and_Renno_2006} who refer to \citet{Lebedev_and_Skalskaya_1962} for detailed calculations of a conducting sphere on a conducting plane. We consider our simplification suitable here.
With the parameters $r= 1 \, \rm mm$, $d = 4.8 \, \rm cm$ and $U = 8 \, \rm kV$, this results in a maximum charge $q_{max} \sim 1 \cdot 10^8 \, \rm e$. This fits the maximum charge measured after recharge and ejection; any particle bound stronger cannot be lifted.

However, as grains are lifted as soon as the electrostatic force equals the sticking force the charge will usually not reach its maximum. In general, it will be set by balancing the electrostatic and the adhesive force 
\begin{equation}
q \cdot E = 3\pi \cdot \gamma \cdot r_d \cdot N.
\end{equation}

In this case, the dust particle number-size distribution peaks at about $r_d = 1 \, \rm \mu m$. It has to be kept in mind that due to grain irregularities, the curvature of a dust grain in contact as well as the sticking force of individual contacts might vary.

With measurements or estimates of all values, we can evaluate whether the charge per contacts are reasonable for a repulsion model. We therefore calculate
\begin{equation}
\frac{q}{N} = \frac{3\pi \cdot \gamma \cdot r_d \cdot d}{U}
\label{qN}
\end{equation}
which gives $q/N = 7 \cdot 10^6 \, \rm e$ as charge to repel aggregates bound by a single contact. It is a reasonable assumption that aggregates only have a small number of contacts so that they are all ejected, eventually. This charge value is therefore in perfect agreement with the idea that dust aggregates charge until the repelling force is strong enough to break the few bonds holding the aggregate to the electrode. 
The charge per contact value is about a factor 10 smaller than the maximum charge which a mm-aggregate can hold. This implies that all aggregates with less than 10 contacts are ejected.

This model also explains why not all aggregates have the same absolute charge after ejection. Depending on the number of contacts that the aggregate made upon impact and the nature of the few individual dust grains participating, the amount of charge necessary to eject aggregates varies. And as each sticking event always results in new combinations of contacts, individual aggregates also show variations in their charges during subsequent recharging events.

While the maximum charge scales with aggregate size, we assume that the sticking force remains the same, even though the exact impact of the aggregates' size on the adhesive force cannot easily be quantified. We observed that bigger aggregates do not necessarily have more contacts as they are more prone to move or sway during the recharging process. This would explain how aggregates way smaller than 1\,mm do not reach the charge necessary to overcome adhesion.

Size also seems to shift the time interval needed between contact and lift-off, i.e. smaller aggregates are ejected at later times in comparison to the larger aggregates. While we did not set up a quantitative charging model regarding the expected contact variety, aggregates will likely charge like little capacitors, i.e. exponentially approaching their final charge state if adhesion is strong enough. In any case, the absolute charge gained in a certain time interval is larger on larger grains. So this is also in agreement to the observations.

This repulsion model is rather simplified, especially on the adhesion side. The video data show that some of the aggregates slightly sway while sitting on the electrodes. This suggests that the number of contacts during recharging and, thereby, the adhesive force is not constant. Besides, when aggregates first come in contact with an electrode they often have enough momentum to slide across the copper surface before coming to a halt, sometimes losing material in the process. Water that is clinging to the aggregate surface or the electrode also affects the stickiness of the particles \citep{Steinpilz2019, Pillich2021}. All this would change or add to the adhesive force at hand, complicating eq. \ref{qN} and changing the charge that needs to be accumulated in order to overcome sticking forces. However, the results of our experiment can be adequately explained by this simplified model and the idea that dust aggregates are moderately electrically conductive. In fact, besides the unknown details of the exact contact number, sizes and ohmic resistance and without taking dynamic processes into account, the lifting force on the dust aggregates can be readily estimated.

\subsection{Conducters versus insulators}

The small amount of pre-charge is likely gained during the injection process as aggregates should have been discharged during the waiting time until launch, which is way longer than the fraction of a second needed to recharge at the electrodes. The bouncing collisions show that charge on this order of magnitude can be transferred in a single collision, so interaction during injection seems the most likely candidate to set the pre-charge.
Otherwise, the maximum charges are set by conduction and, on maximum, are on the order of $10^8 ~ \rm e$.

Looking into the charges that are accumulated over a certain time, we can determine the conductivity of the individual aggregates. Here, we observe values of $\sigma = 10^{-13} \ ... \ 10^{-11} \ \rm S/m$. This is similar to observations of \citet{SaintAmant1970}. They investigated basaltic dust samples with a filling factor comparable to our aggregates and observed conductivities of $\sigma = 10^{-13} \ ... \ 10^{-15} \ \rm S/m$ in vacuum and at room temperature. Those values are not very high, but within a strong electric field it is sufficient to permit current flow through the sample.

As much as conductivity of the dust aggregates is the natural explanation in these experiments, an insulating nature of other samples was evident in earlier experiments on similar size, but using monolithic basalt and glass beads \citep{Jungmann2021x,Jungmann2022x}. In those experiments with the same setup, particles were shown to have permanent dipole moments for long times, just to mention one piece of evidence that rules out conductivity \citep{Steinpilz2020b}. This was confirmed by \citet{Onyeagusi2022}. 
The charging model by \citet{Wang2016} explaining electrostatic repulsion from atmosphereless bodies also assumes non-conductive dust grains to collect various charge patches in contrast to conductive grains.

Some parameters in our current setting were different compared to experiments by \citet{Jungmann2022}, a dry $CO_2$ atmosphere in the earlier experiments for one. So, considering either conductive or non-conductive grains is not a contradiction, but might be set by the environment and particle sample. As we conducted the experiment under ambient pressure, water might play a significant role. \citet{Strangway1972}, who investigated electric properties of dust samples including basalt and lunar dust, found that the water content at ambient pressure would increase the conductivity about 4 orders of magnitude compared to the observed values at 10$^{-7}$~mbar. Using the moisture analyzer scale PCE-MB 60C, we determined a water content of at least 12\,\% in our sample.
It is intrinsically difficult to remove water from the pore space of dust aggregates and it is likely that grains were just humid enough to grant conductivity. An intrinsic conductivity of the material is possible as well. At this stage, we cannot pinpoint the cause of the conductivity.

It is interesting to note that the insulating grains in the earlier experiments by \citet{Jungmann2022} had charges on the order up to $10^8 ~ \rm e$ as well. They had both polarities. Therefore, not all grains would be lifted, but with respect to e.g. friction velocity thresholds of lifted particles, we caution that non-conductive grains might also work to reduce lifting thresholds. This brings us to applications on planetary surfaces.

\section{Planetary applications}

In an electrified setting e.g. during eolian transport, electric fields can have an influence on particle lifting. Gravity also has to be compensated, but in any case, the electrostatic lifting force can easily be applied as done by \citet{Kok_and_Renno_2006}, for example. Here, we quantified the lifting and restraining forces for dust aggregates.
It is a prerequisite, however, that soil and individual grains are conductive in agreement with work by \citet{Holstein2012}. 

It remains an open question, whether strong electric fields can be generated in a conductive setting in the first place. We cannot give a definite answer as of now. The grains in our experiment are not highly pre-charged. It remains to be seen, if that suffices to produce high electric fields. 

Also in view of earlier experiments with glass beads, not all grains might be conductive, especially in arid regions with eolian activity. 
It is curious, however, that the net charges are on the same order of magnitude. 
In that case, grains might also be lifted, albeit due to the fact that some of them are already highly charged. This is a different mechanism not working by conduction, but it would yield similar results as dry tribocharged mm-grains can hold similar amounts of charge as conduction would yield. Such grains would not stick first, but continue to bounce.

The experiments were carried out at an ambient pressure of 1 bar. An atmosphere influences the process by several ways. It might come with a humidity which influences the conductivity of the grains and the sticking forces. On a planet, an atmosphere would also shield from cosmic radiation and in parts from UV radiation, which could provide another charging mechanism, establishing an electric field. Also, atmosphereless bodies typically have less gravity.
In any case, \citet{Kimura2022} and \citet{Hirata2022} point out that there are also strong electric fields on the surface of atmosphereless bodies due to photoeffect or plasma charging. So electric repulsion might be present as well. They require very low sticking forces though to explain a lift of small dust grains. That would be a different setup then, which our experiments do not directly mimic.
\section{Conclusions}

We find in drop tower experiments that dust aggregates on an electrode recharge in an electric field and are repelled as soon as the Coulomb force outweighs the sticking force. This way, large mm-size aggregates are easily lifted in a fraction of a second after sticking to the electrode, while small aggregates cannot gain enough charge for lift-off.
In agreement to earlier works by \citet{Kok_and_Renno_2006} and \citet{Holstein2012}, conduction is one viable mechanism in aeolian transport to reduce the necessary gas drag for lift, but if adhesion is lower it might also provide electrostatic repulsion from atmosphereless bodies \citep{Kimura2022, Hirata2022}.

\section{Acknowledgments}
This project is supported by DLR Space Administration with funds provided by the Federal Ministry for Economic Affairs and Climate Action (BMWK) under grant number DLR 50 WM 2142. We appreciate the helpful reviews of two anonymous referees.

\bibliography{bib}
\bibliographystyle{aasjournal}

\end{document}